\def\editing{}

\documentclass[sigplan,screen]{acmart}

\copyrightyear{2024}
\acmYear{2024}
\setcopyright{acmlicensed}\acmConference[ASPLOS '24]{29th ACM International Conference on Architectural Support for Programming Languages and Operating Systems, Volume 3}{April 27-May 1, 2024}{La Jolla, CA, USA}
\acmBooktitle{29th ACM International Conference on Architectural Support for Programming Languages and Operating Systems, Volume 3 (ASPLOS '24), April 27-May 1, 2024, La Jolla, CA, USA}
\acmDOI{10.1145/3620666.3651326}
\acmISBN{979-8-4007-0386-7/24/04}

\usepackage{float}
\usepackage{listings}
\usepackage{xcolor}
\usepackage{color}
\usepackage{soul}
\usepackage{booktabs}
\usepackage{graphicx}
\usepackage{multicol}
\usepackage{enumitem}
\usepackage{listings}
\usepackage[export]{adjustbox}
\usepackage[most]{tcolorbox}
\usepackage[absolute,overlay]{textpos}
\usepackage{algorithm}
\usepackage[noend]{algpseudocode}

\definecolor{codegreen}{rgb}{0.04314,0.6745,0.07843}
\definecolor{codegray}{rgb}{0.7059,0.6863,0.702}
\definecolor{codered}{rgb}{0.5373,0.02745,0.06275}
\definecolor{codeblue}{rgb}{0.071,0.0258,0.9882}
\definecolor{codepurple}{rgb}{0.6,0.02745,0.5961}
\definecolor{codeblack}{rgb}{0,0,0}
\definecolor{codegrey}{rgb}{0.5,0.5,0.5}
\definecolor{backcolour}{rgb}{0.95,0.95,0.92}
\newcommand{\secref}[1]{\S\ref{#1}}

\newcommand{\figref}[1]{Figure~\ref{#1}}

\newcommand{\algoref}[1]{Algorithm~\ref{#1}}

\algrenewcommand\algorithmicindent{1.0em}%

\newcounter{remark}[section]

   \def\editingmarks{}

\ifx\editingmarks\undefined
\newcommand{\myremark}[3]{}
\else
\newcommand{\myremark}[3]{
\refstepcounter{remark}
\[
\left\{
\sf
\parbox{0.8\columnwidth}
{
{\bf {#1}'s remark~\arabic{section}.\arabic{remark}:}
{#3}
}
\right.
\]
\marginpar{\bf {#2}.~\arabic{section}.\arabic{remark}}
}
\fi

\newtcolorbox{notebox}[3][]{
  enhanced,
  colback=#3!10,
  colframe=#3,
  fonttitle=\bfseries,
  title=#2,
  #1
}

\newcommand{\note}[1]{
\ifx\editing\undefined
\else
	\begin{notebox}{Note}{gray}
			#1
	\end{notebox}
\fi
}

\newtcolorbox{assignbox}[3][]{
  enhanced,
  colback=#3!10,
  colframe=#3,
  title=#2,
	#1
}

\newcommand{\assign}[1]{
\ifx\editing\undefined
\else
	\colorbox{pink}{Assigned to #1}
\fi
}

\newtcolorbox{remarkbox}[3][]{
  enhanced,
  colback=#3!10,
  colframe=#3,
  fonttitle=\bfseries,
  boxed title style={
    colframe=#3,
  	colback=#3!10,
  },
  title=#2,
  #1
}

\newcommand{\alaska}[0]{\textsc{Alaska}}
\newcommand{\anchorage}[0]{\textsc{Anchorage}}

\newlist{todolist}{itemize}{2}
\setlist[todolist]{label=$\square$}
\usepackage{pifont}

\DeclareGraphicsExtensions{.pdf,.png,.jpg,.svg}

\begin{document}

\def\point#1{{\bf #1}}

\title{Getting a Handle on Unmanaged Memory}

\date{}

\thispagestyle{empty}

\author{Nick Wanninger}
\affiliation{
  \institution{Northwestern University}
  \country{Evanston, IL, USA}
}
\email{ncw@u.northwestern.edu}

\author{Tommy McMichen}
\affiliation{
  \institution{Northwestern University}
  \country{Evanston, IL, USA}
}

\author{Simone Campanoni}
\affiliation{
  \institution{Northwestern University}
  \country{Evanston, IL, USA}
}

\author{Peter Dinda}
\affiliation{
  \institution{Northwestern University}
  \country{Evanston, IL, USA}
}
\email{pdinda@northwestern.edu}

\begin{abstract}
The inability to relocate objects in unmanaged languages brings with it a menagerie of problems.
Perhaps the most impactful is memory fragmentation, which has long plagued applications such as databases and web servers.
These issues either fester or require Herculean programmer effort to address on a per-application basis because, in general, {\em heap objects cannot be moved in unmanaged languages}.  
In contrast, managed languages like C\#\ cleanly address fragmentation through the use of compacting garbage collection techniques built upon heap object movement.
In this work, we bridge this gap between unmanaged and managed languages through the use of handles, a level of indirection allowing heap object movement.
Handles open the door to seamlessly employing runtime features from managed languages in {\em existing, unmodified code} written in unmanaged languages.
We describe a new compiler and runtime system, \alaska{}, that acts as a drop-in replacement for \texttt{malloc}.
{\em Without any programmer effort}, the \alaska{} compiler transforms pointer-based code to utilize handles, with optimizations to minimize performance impact.
A codesigned runtime system manages this new level of indirection and exploits heap object movement via an extensible service interface.
We investigate the overheads of \alaska{} on large benchmarks and applications spanning multiple domains.
To show the power and extensibility of handles, we use \alaska{} to eliminate fragmentation on the heap through defragmentation, reducing memory usage by up to 40\% in Redis.
\end{abstract}

\maketitle{}

\section{Introduction} \label{sec:introduction}

A vast amount of code has been written in unmanaged languages such as C, C++, Fortran, and others. 
The billions of lines of code written in these languages are not going anywhere.
C, C++, and Fortran are the 2nd, 3rd, and 14th hottest languages in the TIOBE 2022 index \cite{TIOBE-2022}.
A major draw of these unmanaged languages, particularly C, is the direct control over memory management granted to developers.
Even in application code, this control is nearly at the machine level using raw memory addresses.
The programmer can carefully control object placement and lifetime, even specifying the representation of a pointer.
For example, one can freely encode type information into unused address bits, store pointers as integers, multiplex pointers in an XOR linked list, or even send and receive pointers over a network.

This power brings with it the potential for bugs and security vulnerabilities, almost as if the programmer {\em were} working at the machine level.
In this work, we focus on the {\em limitations} of memory management in unmanaged languages compared to managed ones.
In particular, managed languages have their own clear advantage: heap objects can be straightforwardly moved by the runtime system.

The intentionally designed Wild West semantics of pointers in unmanaged languages makes it {\em virtually impossible} for the language runtime system to move an object.
In a managed language, it is possible to algorithmically identify all pointers to an object, and thus update them when the object is moved.
To do this in an unmanaged language would require understanding pointer semantics {\em as they are for that specific program.}
For example, a memory manager would need to be able to update through XOR list encodings, union types, and address masking.
Generally finding direct pointers is challenging enough.
Generally finding bespoke-encoded pointers is beyond the pale.

The inability of an unmanaged language runtime to easily move heap objects makes their initial placement on the heap a matter of great importance.
This has led to a wide range of heuristics and compiler-driven allocation strategies \cite{LATTNER-AUTOMATIC-2005}.
If the initial placement is wrong, it can have cascading consequences throughout the lifetime of the badly placed object.
One such consequence is external fragmentation---when free space is scattered in small unusable chunks---which often results in using far more physical memory to run the program over the long term than would have been possible with an oracle placement.

To complicate matters further, heap fragmentation often occurs in phases, and thus heap object placement decisions made in one phase of the program could very well induce fragmentation in another phase, even if a perfect oracle was available.
It has been shown that {\em any} allocation strategy that is not free to relocate objects will suffer from fragmentation \cite{ROBSON-1977}.
This reality forces the programmer to act.

The first option is to adopt the ostrich algorithm\footnote{To stick one's head in the sand and ignore the problem.} and allow fragmentation to fester, punting the resulting memory usage issue down the road.
This forces the user to deal with unexpected results such as application restarts or the kernel's out-of-memory killer.
Worse even, the user may need to {\em pay for more memory} on a cloud hosting service just to run their fragmented application.

The second option is for the programmer to attack the fragmentation problem head-on with an ad-hoc defragmentation strategy.
For example, the programmers behind Redis, a popular in-memory key-value database system, have added \texttt{activedefrag} in version 4.0 \cite{REDIS-WEBSITE,REDIS-ACTIVEDEFRAG}.
Because unmanaged pointers are so hairy and have tendrils everywhere, \texttt{activedefrag} required the addition of thousands of lines of code to handle all edge cases, mind you, just the ones in Redis.
Through this considerable engineering effort, \texttt{activedefrag} enables Redis to tackle fragmentation through modifications to the allocator (jemalloc), polling the {\em fragmentation ratio} (RSS over heap usage) of the program once a second to decide if it should defragment.
This approach tends to reduce fragmentation substantially.
Regrettably, the benefits of such a hand-crafted system cannot be transferred to other applications without a great deal of reengineering to handle new data structures and pointer semantics.

In contrast, programmers in higher-level managed languages such as Java or C\# have long enjoyed {\em transparent} object relocation and heap compaction, allowing them to focus on application features rather than worrying about where objects are located.
This is possible due to the guarantees about how pointers are used and stored---having very strict pointer semantics throughout the language.
This vastly simplifies the operation of updating references when a heap object is moved, and the runtimes of these languages often utilize read/write barriers, safe pointing, and
forwarding pointers to increase efficiency \cite{ZORN-TECH-1990,JOHNSON-READ-BARRIER-1991,PIRINEN-BARRIER-1999,PIZLO:2010:SCHISM}.

Recent work addresses the general problem of fragmentation in unmanaged languages {\em without} object mobility.
Mesh \cite{POWERS-MESH-2019} leverages virtual memory to map multiple heap extents in the virtual address space to the same extent in the physical address space.
The allocator is co-designed with this capability to locate objects (permanently) in the virtual heap extents such that there are no collisions in the physical address space and that theoretical guarantees can be provided about the packing of the objects into the physical address space.
In effect, the virtual heap is now much larger with considerable sparsity, but the physical heap is dense.

In this work, we argue that it is possible to add the generic capability to move heap objects to the Wild West of unmanaged languages through the use of {\em handles}.
This generic object movement capability in turn allows for the design of effective, application-independent defragmentation, and opens the door for other services.
Adding this generic ability can be done in a manner transparent to the programmer, even if they make use of complex, custom pointer semantics---the hallmark of unmanaged languages.

The concept of handles, elaborated on in~\secref{sec:handles}, dates back to at least the early days of personal computers, prior to the inclusion of virtual memory.
In handle-based memory management, the allocator doles out opaque handles instead of raw pointers to heap allocations.
Handles must be pinned by the application before use and then unpinned afterward.
Pinning provides a (current) raw pointer to the object.
If a handle is unpinned, its corresponding heap object can be moved.
In classic Windows and MacOS, all applications shared a single heap managed in this manner, which was regularly compacted.
Unfortunately, handles were a directly visible feature of these systems and could be easily misused by programmers, bringing down the whole house of cards.

\begin{figure}[t]
	\includegraphics[width=1\linewidth]{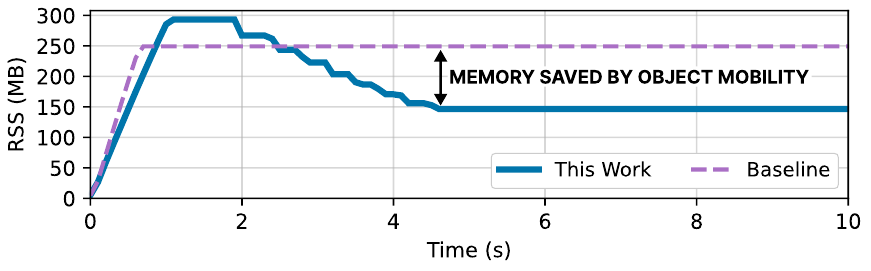}
	\caption{Through object mobility, \alaska{} can save considerable memory---up to 40\% in Redis.}
	\label{fig:sneak-peek}
\end{figure}

Our work has three key differences.
First, we show how the Herculean effort of using handles correctly and efficiently is avoided entirely through compiler analysis, transformation and optimizations.
Second, our method makes handles {\em entirely transparent.}
Programmers can simply write pointer-based code with whatever specialized semantics fit their fancy, with handle-based code running under the hood.
Even better, the programmer cannot possibly get handles wrong---because they don't even see them---lifting the Sisyphean burden of handle-based code maintenance from their shoulders.
This transparency also means our work can be directly applied to {\em existing, unmodified applications}.
Finally, we show how a careful compiler/runtime co-design can implement transparent handles with a geomean overhead of 10\%---a surprising result as handles add a layer of indirection.

\alaska{} and the test suite to reproduce results are publicly available.
See the Artifact Appendix for more information.

\clearpage

\point{This paper makes the following contributions:}
\begin{itemize}[leftmargin=*]
	\item We make the case for revisiting handle-based memory management
		as a mechanism for bringing general-purpose heap object mobility
		to unmanaged languages. (\secref{sec:handles})

	\item We describe the design of a  handle-based memory management
		system that relies on compiler/runtime co-design to {\em automate} the
		use of handles, and to make the use of handles {\em transparent} to the
		programmer (\secref{sec:design}). 
		We implement \alaska{}, an extensible proof-of-concept for C/C++ (\secref{sec:impl}).

	\item We evaluate the overhead of \alaska{} on a number of well-established C and C++
		benchmarks and applications with {\em zero code modifications}.
		The 10\% overhead (geomean) suggests the practicality of automatic transparent handle-based memory management (\secref{sec:eval}).

	\item We implement \anchorage{}, a memory allocator that enables dynamic run time heap defragmentation in unmodified C and C++ applications on top of \alaska{} (\secref{sec:anchorage-impl}).
	\item We show that \anchorage{} reduces memory usage in Redis by up to 40\% through defragmentation, on par with activedefrag: a bespoke tool built specifically for Redis (\secref{sec:redis-defrag}, also \figref{fig:sneak-peek}).
\end{itemize}

\section{Handle-Based Memory Management}
\label{sec:handles}
We now describe classic handle-based memory management, whose motivation to move heap objects mirrors our own.
Unfortunately, in classic systems, this capability came with high programmer effort and, {\em because it was manual}, the possibility for disastrous programmer error.

\subsection{Manual Handles of Yore}
\label{sec:macos}

Handles were originally a solution to a hardware problem of their time, namely that early Motorola and Intel CPUs did not provide hardware MMUs.
Without the hardware underpinnings of virtual memory, PC operating systems, such as early versions of classic MacOS and Windows, required an alternative.
Classic MacOS had only {\em one (physical) address space}, with a single global heap shared between applications \cite{APPLE-VOLUME-2}.
Thus, if one application created fragmentation on the heap, the entire system felt the consequences---{\em including the kernel}.
This was especially problematic given the memory restrictions of the time: the original Mac had only 128 KiB of RAM.
Handles enabled these systems to more efficiently use their already limited memory through defragmentation.

\begin{figure}[b]
	\includegraphics[width=1\linewidth]{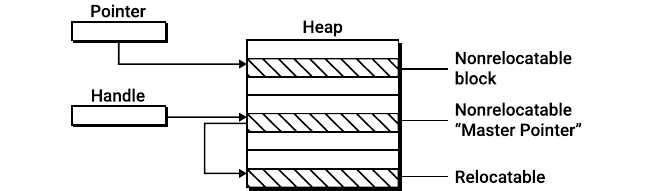}
	\caption{MacOS handles pointed to a relocatable block.}
	\label{fig:mac-heap-layout}
\end{figure}

In this system, an allocated block in the global heap may either be {\em relocatable} or {\em nonrelocatable}.
Relocatable blocks could be moved within the heap to make space through defragmentation so long as all users followed the contract: these blocks may be either {\em locked},\footnote{We adopt the classic terminology here, but it is important to keep in mind that handle locking is unrelated to concurrency control. The modern terminology is {\em pin}.} preventing movement, or {\em unlocked}, permitting movement.\footnote{
Beyond movement, blocks could also be marked ``purgeable'' or ``swappable'', allowing the block to be discarded or moved to disk to make space.}
When a relocatable block was allocated, the memory manager would create and maintain a single nonrelocatable {\em master pointer} to that block.
A pointer to this master pointer was returned to the programmer, being a handle that must be translated before use.
This relationship is shown in~\figref{fig:mac-heap-layout}.

Simply put, handles provided a single level of indirection between logical references to an application's heap data and the actual backing storage.
This indirection {\em allowed objects to be freely moved on the heap by updating only a single reference}: the master pointer.

\subsection{Manual Handles Are Cumbersome}

While handles gave the system a great deal of control with regard to object placement, manual handles were considered a nightmare to program for several reasons.

\begin{figure}[h]
\begin{lstlisting}[language=c]
int **handle = NewHandle(...);
HLock(handle); // Lock/pin
int *ptr = *handle; // Translate 
*ptr = 42; // Use memory
HUnlock(handle); // Unlock/unpin
\end{lstlisting}
\caption{Manual handles required significant effort with uncomposable changes that permeate the application.}
\label{fig:code-handles}
\end{figure}

\figref{fig:code-handles} shows an example use of MacOS handles (translated from the original Pascal to C).
Manually adding handles to a program required significant effort---the majority of the lines shown are concerned with their management.
Handles are allocated with the \texttt{NewHandle} function.
In order to safely access the data behind the handle the user must first call \texttt{HLock}, setting a bit in the handle to indicate it has been locked/pinned.
This allows the programmer to safely access the backing memory of the handle without concern for object movement.
When done accessing the object, programmers must call \texttt{HUnlock}.

With manual handles, programmers were required to consider an additional lifetime---that of their pins---which came with tradeoffs.
Being overly cautious with pins (i.e., surrounding individual memory accesses with pin/unpin) is easier to understand, but produces terrible performance from the additional memory updates needed to pin, unpin and translate handles.
Conversely, pinning across large intervals (e.g., an entire loop or function) reduces the performance issue, but may result in many nonrelocatable blocks, limiting the effectiveness of defragmentation when needed.

In addition to thinking about when to free memory---avoiding double frees, use-after-free, and memory leaks---the programmer must also be concerned with the effects of unpinning.
When an object is unpinned, it {\em can be moved}; unpinning too early could lead to the object being moved without the application's knowledge.
The programmer also had to ensure they were not unpinning a handle multiple times, and that they unpin before the handle goes out of scope, lest it be pinned forever.

Finally, manual handle-based memory management puts the onus of correctness entirely on the programmer.
Worse, on a modern system, programmers would need to consider correctness in the presence of preemptive threads and signals, neither of which existed in classic systems.

\section{Design of the \alaska{} System}
\label{sec:design}

We assert that automatic handle-based memory management holds the key to achieving object mobility in unmanaged languages, empowering the runtime to freely relocate heap objects and more.
Guided by this, we present the design of \alaska{}---a compiler and runtime system whose goal is to achieve {\em automatic transparent handles}.
Handles managed by \alaska{} gracefully coexist with pointers.
This seamless integration ensures that programmers can unknowingly wield handles with confidence just as they would traditional pointers. 
This also means that {\bf entirely unmodified applications can now automatically leverage handles}.

\subsection{Design Goals}
\label{sec:design-transparency}
\label{sec:goals}

A key objective of \alaska{} is to enable the movement of heap objects using handles with zero additional programmer effort.
Such a system would allow the billions of lines of code written in unmanaged languages to benefit from managed techniques currently held out of reach.

\point{Handles and pointers coexist.}
In order to support all existing applications written in unmanaged languages, raw pointers and handles must coexist.
This is necessary because a function written to accept a pointer must behave in the same way regardless of it being passed a pointer or a handle.
Thus, a variable can hold either a raw pointer---as it would originally---or a handle.
As far as the programmer is aware, then, handles are just pointers with no additional semantics.

\point{Handles have the same semantics as pointers.}
To maintain the original application's functionality, handles {\em must} have the same programmer-visible semantics as the pointers they replace.
This means that handles must support the majority of pointer encoding and multiplexing techniques outlined in \secref{sec:introduction} that keen programmers may employ.
So long as the application does not violate the assumptions outlined in \secref{sec:assumptions}, no changes must be made in most applications to use handles. As our implementation is IR-level, this is probably also true for other languages but is untested.

\point{The compiler does translation for you.}
The most important component of \alaska{} is the compiler, which manages the translation and tracking of handles for the programmer.
Unlike the handles of old (\secref{sec:handles}) the programmer should not need to modify their application in any way to take advantage of the benefits of handles.
This is achieved via the \alaska{} compiler, which automatically translates handles and ensures they are both correct and optimized.

As discussed in \secref{sec:handles}, small pin intervals (individual load/stores) grant the most freedom for memory management, but can incur a worrying performance overhead.
Conversely, large pin intervals (loops or functions) reduce the performance impact of pinning, but reduce the degrees of freedom available to the memory manager.
Through the compiler, the tradeoff between small and large intervals can be explored automatically, leaving one less dragon looming over the development process.
We present one such optimization to hoist pins outside of loops when beneficial (\secref{sec:impl-compiler}).

\point{The runtime efficiently tracks pins for you.}
\alaska{}'s runtime automatically manages the tracking of which handles are in use and which are not.
However, a runtime system that na\"{i}vely records pins as in \figref{fig:mac-heap-layout} would perform poorly today in multithreaded environments.
Concurrent updates to the shared structures using atomics would lead to contention across the machine, especially when handle pins occur at a high rate.
The runtime system must correctly handle such concurrent access without significantly affecting other concurrency control logic within the program itself.

\subsection{Assumptions} \label{sec:assumptions}

While handles in \alaska{} do not affect the programmer-visible semantics of pointers, \alaska{} imposes a limited set of restrictions on their usage, to enable a more performant implementation.
We argue that these assumptions are not fundamental to transparent handles, but are rather design decisions specific to our implementation.

We assume that a program will not access memory outside the bounds of the allocation returned by the allocator.
If the assumption is not true, \alaska{} can potentially translate the wrong handle, or generate an out-of-bounds access.
This is the same assumption made by LLVM's memory model \cite{LLVM} and can be considered a requirement for any meaningful memory transformation.
Similarly, the optimizations in our compiler rely on the application not breaking strict aliasing assumptions (casting a pointer of one type to a pointer of an incompatible type), which GCC and Perlbench from SPEC CPU, unfortunately, do \cite{SPEC-GCC,SPEC-PERLBENCH}.

We also assume that programs do not rely on the bit representation of a pointer outside of the alignment guarantees specified by \verb.malloc.. 
With this, implementations are permitted to utilize lower address bits for their own purposes. They are not, however, allowed to assume the top bits of a pointer are free to use (as they are on x86).
This unfortunately rules out the use of NaN-boxed handles, as such systems often rely on virtual addresses being only 48 bits.

The final assumption of the system relates to external libraries: we assume that either all library code is subject to our compilation, or that it does not store pointers to the backing memory nor free that memory.
More on this assumption and our approach for handling it are discussed in \secref{sec:compiler-escape-impl}.

\subsection{Efficient Handle Translation}
\label{sec:trans}

\alaska{} is built as a pure software solution, meant to run on commodity hardware.
Without hardware acceleration, as is the case for virtual memory, {\em translation} from handles to pointers must be performed in software using standard instructions.
A poor design could be catastrophic for performance.
As such, the design of the handle representation leans upon the understanding of how pins will be lowered into the ISA, minimizing the number of additional loads.\footnote{This means we can't use a traditional radix tree as seen in virtual memory, as it would quadruple memory accesses.}
We specifically consider x64, ARMv8.3, and RISC-V 64-bit architectures in our design---but only evaluate x64 for brevity.

\begin{figure}
	\includegraphics[width=0.9\linewidth]{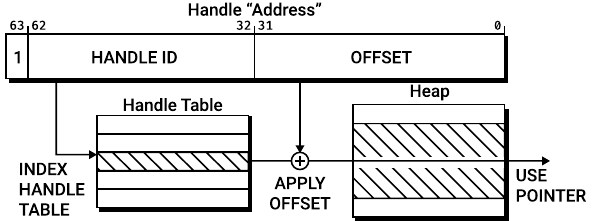}
	\caption{Handles in \alaska{} are pointers (hardware-level addresses) with the top bit set.
  The Handle ID is used to index into the handle table, and the offset is added to the resulting address.}
	\label{fig:handle-repr}
\end{figure}

Given these requirements, we chose the bit representation for handles as shown in \figref{fig:handle-repr}.
With this representation, a handle in \alaska{} is differentiated from a pointer by the top bit---being set to $1$ for a handle, or $0$ for a pointer.\footnote{Note that on all architectures considered, this will result in a handle being accidentally used as an address to fault.}

If the top bit is set, bits 32 to 62 represent the {\em handle ID}, which acts as an offset into the {\em handle table} data structure.
Conceptually, this is the same way a virtual address is broken into offsets into the various levels of the page table hierarchy, but with only a single level.
Each allocation in the system has a unique handle ID, and the design effectively limits the number of active handles in the system to $2^{31}$.
The decision to restrict the number of bits in the handle ID is influenced by many architectures' ability to efficiently truncate values to 32 bits, making it a practical choice.
The handle representation's lowest 32 bits are used to denote an {\em offset} into the object, capping the maximum object size to 4GiB.
We contend that this limitation is not a significant concern, as relocating objects $\ge4$KiB can be more efficiently handled by paging.

These design decisions allow us to implement \alaska{}'s handle translation logic in only 6 instructions on x64 (\figref{fig:code-translate}).
We investigate the overhead of these additional instructions in \secref{sec:eval}.

\begin{figure}[t]
\begin{lstlisting}[language={[x86masm]Assembler}]
  cmp  -2,%rdi     ; Handle check
  jg   skip        ; Not a handle
  mov  %rdi,%rax
  shr  0x1d,%rax   ; Extract handle ID
  mov  %edi,%edi   ; Truncate offset
  add  (%rax),%rdi ; HTE Load + offset
skip:
  mov  (%rdi),%rdx ; Perform access
\end{lstlisting}
\caption{x64 instructions to perform a handle translation.}
\label{fig:code-translate}
\end{figure}

\subsection{Efficiently Tracking Pinned Handles}
\label{sec:tracking-design}

On top of managing translations, a key functionality of \alaska{} is to record
which objects are in use and which are not, a process referred to as ``pinning''.
This is required because \alaska{}'s compiler transformation will leave references to the raw \textit{backing memory} in CPU registers or spilled on the stack.
As such, pinned handles are viewed as a constraint in the runtime system, and cannot be moved.

\subsection{Extensible Service Interface}
\label{sec:design-service-interface}

Beyond the core functionality of the \alaska{} runtime, an extensible interface allows for the addition of {\em services}: pluggable and configurable libraries that manage the allocation and movement of objects.
The separation of these services from the core runtime enables exploration into different techniques that can make use of the object mobility provided by handles (\secref{sec:discussion}).
In this paper we developed one such service, \anchorage{}, to perform defragmentation.
\anchorage{} provides \alaska{} with an allocator and barrier routine designed to perform heap defragmentation.
We describe its implementation details in \secref{sec:anchorage-impl}.

\section{Implementation of \alaska{}}
\label{sec:impl}

The \alaska{} System comprises three logical components: a compiler that
transparently automates the use of handles in unmodified pointer-based
code, a core runtime that manages the handle table and tracking,
and a generic interface to allow the construction of services such as
\anchorage{}.

\subsection{The \alaska{} Compiler}
\label{sec:impl-compiler}

\alaska{}'s compiler transforms programs to use handles by replacing allocation routines (\secref{sec:compiler-conversion-impl}), optimizing the translation of handles (\secref{sec:compiler-translation-impl}), and inserting tracking for pinned handles (\secref{sec:tracking-impl}).
It also maintains correctness with external libraries through an {\em escape pass} (\secref{sec:compiler-escape-impl}).
We operate on the LLVM IR \cite{LLVM} and use abstractions from NOELLE \cite{NOELLE}.

\begin{figure}
  \includegraphics[width=1\linewidth]{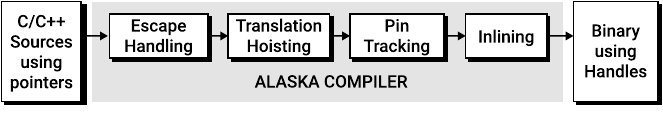}
  \caption{\alaska{}'s transformation pipeline.}
  \label{fig:compile}
\end{figure}

\subsubsection{Replacing malloc with halloc}
\label{sec:compiler-conversion-impl}
Conversion of memory allocations to handle allocations is straightforward.
Each call to \texttt{malloc} and \texttt{free} is transformed by the compiler to a call to \texttt{halloc} and \texttt{hfree}---their \alaska{} counterparts.
This is also the case for proxy functions such as \texttt{calloc} and \texttt{realloc}.
We perform this replacement in the compiler and not the linker to avoid introducing handles into code that is not visible to \alaska{}.
This behavior can be disabled to allow the programmer to decide which allocations are made with \texttt{malloc} and which with \texttt{halloc}.
In our evaluation, we force handles on all allocations through \texttt{malloc}.

\subsubsection{Automatically Translating Handles}
\label{sec:compiler-translation-impl}
The translation insertion compiler pass transforms the program so that each instruction that \emph{may} access a handle-allocated object \emph{must} perform a translation beforehand.
To simplify this transformation, \alaska{}'s translation function has different functionalities depending on the incoming value.
If the incoming value is a pointer, the translation function simply does nothing and returns the pointer.
If it is a handle, the function performs the translation as described in \figref{fig:handle-repr}.

A na\"{i}ve implementation of translation insertion would be to translate immediately before each memory access in the program.
However, this would incur a large runtime overhead: a bit test for every memory access in the best case and a load from the handle table in the worst case.
To prevent this, we present \algoref{alg:translation-insertion}, which analyzes and transforms an LLVM program to minimize the number of dynamic translations.
The shorthand $\mathit{ptr}(inst)$ represents either the address of load/stores, results of a \texttt{translate}, or the operand of transient values (i.e. $\phi$ and \texttt{getelementptr}).

\noindent

\begin{algorithm}[]
  \caption{The translation insertion algorithm.}
  \small
  \noindent
  \begin{algorithmic}[0]

  \Procedure {\small{TranslationInsertion}}{$F: \mathrm{function}$}

    \State $PG \leftarrow \mathrm{pointer\ flow\ graph\ of\ } F$
    \State $PG' \leftarrow \{p \in PG\ |\ \mathrm{incoming}(p) > 0\}$
    \State $DT \leftarrow \mathrm{dominator\ tree\ of\ } F$
    \State $DF \leftarrow \{n \in DT\ |\ n \in PG'\}$ \Comment A dominator forest.
    
    \ForAll {$t \in DF$} \Comment For all trees in forest.
      \State $r \leftarrow \mathrm{root}(t)$
      \State $l \leftarrow \textsc{\small{Translate}}(r, F)$
      \ForAll {$n \in t$}
        \State $\mathit{ptr}(n) \leftarrow l$ \Comment Replace handle with the translation.
      \EndFor
    \EndFor
  \EndProcedure
  \Procedure {\small{Translate}}{$i, F$}
    \State $LT \leftarrow \mathrm{loop\ nesting\ tree\ of}\ F$
    \State $L \leftarrow \{L \in LT\ |\ L\ \mathrm{is\ the\ innermost\ loop\ containing}\ i\}$
    \State $L' \leftarrow \textsc{\small{FindNestingLoop}}(L, LT, i)$
    \State $BB \leftarrow \mathrm{preheader\ of}\ L'$
    \State $t \leftarrow \mathrm{terminator\ of}\ BB$
    \State Insert \texttt{translate(ptr($i$))} immediately before $t$
    \State \Return The result of \texttt{translate(ptr($i$))}
  \EndProcedure
  \Function {\small{FindNestingLoop}}{$L, LT, i$}
    \State $L' \leftarrow \mathrm{parent}(L) \in LT$
    \If {$i \in L'\ \wedge \mathrm{ptr}(i) \notin L'$}
      \Return $L'$
    \ElsIf {$i \notin L'$}
      \Return $L$
    \Else {}
      \Return $\mathrm{\textsc{\small{FindNestingLoop}}}(L', LT, i)$
    \EndIf
  \EndFunction

  \end{algorithmic}
  \label{alg:translation-insertion}
\end{algorithm}

This transformation ensures that each memory access to a handle will operate on the translated pointer to its backing memory as each access is dominated by a \texttt{pin}.
For memory accesses within loops, \textsc{InsertPin} hoists the requisite pins outside of loops when possible.
This pass relies on LLVM's canonical loop form (\texttt{-loop-simplify}).

Following the insertion of translations in the program, the compiler inserts {\em releases} which indicate when the handle is no longer in use.
These translations are only inserted to simplify the implementation of the tracking pass later (\secref{sec:tracking-impl}), and are removed before the program is run.
This is performed with the results of a liveness analysis \cite{DRAGON-BOOK} on each of the translated handles.
For each \texttt{ptr = translate(handle)} inserted into the program, \texttt{release(handle)} calls are inserted immediately at the end of \texttt{ptr}'s lifetime.

\subsubsection{Tracking Pinned Handles} \label{sec:tracking-impl}
Once a handle is translated, it must be pinned to ensure that the backing memory block represented by the handle cannot be relocated for the lifetime of the translation.
This is required because, during the translation's lifetime, raw pointers (i.e., virtual addresses) to the backing memory exist, for example in CPU registers.
In the common case, most applications do not have a large number of pinned handles at any point in time, and thus the runtime is free to move most objects at any time.

An intuitive, but na\"ive implementation of tracking pinned handles might atomically increment a \verb|pin_count|  attached to each handle in the handle table when a translation occurs, and atomically decrement it when the translation's lifetime ends.
When \verb.pin_count.$\ >0$, the handle would be considered pinned and the backing memory would be immobile. 
Unfortunately, these pin/unpin steps would naturally incur undue overhead in applications that exhibit many translations---especially in a multithreaded application, and as core counts grow.  An alternative mechanism is necessary.

In \alaska{}, pinned handles are tracked \textit{privately}, on the call stack, requiring \textit{no atomic instructions}.
For any function that translates handles, we generate code that allocates a single \textit{pin set} in the current stack frame in the function prelude.
At run time, each invocation of the function thus has a private pin set.
The pin set stores the translated handles for the invocation.    
The compilation statically decides the size of each function's pin set so that is large enough to contain at least as many pinned handles as may statically overlap at any point in the function's control flow.
Each \textit{static} translation in the program is allocated a single entry in the function's pin set using a greedy interference graph-based allocation strategy similar to a register allocation algorithm.  At run time, before a handle is translated, the handle is stored in the pin set.

The pin set is stored as an array on the stack, requiring no additional instructions to construct and imparting no additional heap memory usage.
An experimental feature of LLVM's garbage collector infrastructure, StackMaps \cite{llvm:stackmap}, is used to record the array's location relative to the stack or frame pointers (e.g., \texttt{\%rsp} or \texttt{\%rbp}) for each return address in the program.
This information can then be used during run time to walk the current stack with Libunwind \cite{libunwind} to find the corresponding stack of pin sets embedded in it.

\point{Barriers and Pin Set Unification.}
Because each thread tracks its pin sets privately, there must be a mechanism to pause all threads and unify all their extant pin sets into a global pin set before using the information in that set to determine which backing memory blocks are currently immobile. 
We refer to this mechanism as a {\em barrier} to reflect our current implementation: a stop-the-world pause event, during which the runtime is free to relocate objects.
Because LLVM StackMaps are only valid at certain points in the program, the threads cannot be simply interrupted using POSIX signals.
As such, \alaska{} uses safe pointing and polling to ensure that the local pin set is valid at certain points in the program \cite{GCPOINTS:AGESEN,GCAGESENDETLEFS}.

The \alaska{} compiler inserts calls to an LLVM safepoint intrinsic, provided by the StackMaps infrastructure, marking points at which the StackMaps information must be valid.
We place safepoints throughout the program on the back edges of loops, the entry point of certain functions, and before calls to external libraries.
The LLVM backend recognizes these safepoints and emits a {\em patch point} at each corresponding ISA-specific location.
On x64 a patch point is a \texttt{NOP} instruction. 
In the best case, these safepoints have no overhead and produce no register or data cache pressure.
However, as is to be discussed in \secref{sec:eval}, this API is {\em experimental}, and unfortunately does not have zero overhead in all cases.

When the runtime wishes to unify pin sets, a barrier is started and each patch point's \texttt{NOP} instruction is replaced with a \texttt{UD2} instruction, which, when executed, causes an illegal instruction exception that the kernel in turn delivers to the thread's \texttt{SIGILL} handler, which is part of the \alaska{} runtime.
In this signal handler, the StackMaps information is valid, and all pin sets can be parsed safely.

This would be sufficient if all code were transformed by the \alaska{} compiler.
However, as a practical matter, we often need to support external library code, and, at minimum, the system call wrappers need to be considered.
Here, the problem is that there are no safepoints, and, in some cases, blocking in the kernel, e.g., during an I/O operation, can also occur.
Consequently, we might end up waiting for an arbitrary amount of time.
To handle this situation, the barrier mechanism does not simply wait forever for all threads to join.
Instead, if a straggler is detected, that thread is signaled using a POSIX signal, and the handler for that signal effectively contains the safepoint.
This works because there is no handle translation occurring within the external code, and thus no pin sets can exist ``below'' the immediate external call, no matter how deep the call stack is below that point.

Once all threads are in the runtime, they synchronize and determine which handles are pinned and which are not.
The runtime is then free to move backing memory blocks however it sees fit, so long as it obeys the pin status of each handle.
When done, the runtime returns each patch point to its original \texttt{NOP} state, and all threads are resumed.

\subsubsection{External Functions}
\label{sec:compiler-escape-impl}
The \alaska{} compiler assumes a whole-program representation but calls to precompiled libraries such as libc are commonplace.
The most common implementation, \texttt{glibc} \cite{GNU-LIBC} cannot be compiled with \texttt{clang} due to its reliance on GNU extensions.
This leaves the compiler with code that \emph{may} break the assumptions in \secref{sec:assumptions}.
Rather than prohibiting users from using \texttt{glibc}, we handle cases of broken assumptions in turn.

For external functions, the compiler performs \emph{escape handling}.
An \emph{escape} occurs whenever a handle is passed into precompiled code as an argument.
For each escaped handle, the compiler inserts a \texttt{pin} before the call and passes the resulting, raw pointer to the function.
This ensures that the precompiled code will operate correctly.

For cases where assumptions are broken, the compiler must transform the function.
An example of this can be seen in parsing applications, where the size of a token is computed by subtracting the non-handle result from the escaped handle argument of \texttt{strstr}.
This results in integer underflow and often leads to a segmentation fault.
To remedy this, the \alaska{} compiler replaces the external function with its \texttt{musl} \cite{MUSL-LIBC} implementation and transforms it with the rest of the program.
With this solution, \texttt{strstr} operates as expected and the aforementioned issue is avoided.

\subsection{The Core Runtime}
\label{sec:runtime}

\alaska{}'s runtime manages low-level operations, including handle allocation/deallocation, pin/unpin tracking, and the delivery of ``barrier'' operations.
Handle allocation is exposed to the programmer through two functions, \texttt{halloc} and \texttt{hfree}, which mirror the functionality of \texttt{malloc/free} respectively.
These allocation functions are automatically used in place of the system allocation functions in any code that is transformed by the compiler, as mentioned above.

\subsubsection{The Handle Table}
\label{sec:handle-table}

At the center of the core runtime is the {\em handle table}, a metadata structure that enables efficient handle translation.
The {\em handle table} is analogous to the page table in virtual memory systems, albeit with one handle table entry (HTE) per-object instead of per-page.
Unlike a hierarchical page table in x86, ARM, or any other modern system, the handle table in \alaska{} is a single-level table.
This eliminates the need for page-walk-style translations in software, which would impart unreasonable performance overheads.
A simple linear array enables translations to occur with a single load.

The handle table lives in the virtual address space of the user program, and all translations are performed in software using unprivileged instructions.
The table is placed at a specific virtual address so that translations need not mask off the top bit of the handle representation
(\figref{fig:handle-repr}).
To ensure the handle table is placed at the right location and {\em can} use all $2^{31}$ entries it is virtually allocated via \verb.mmap. in its entirety at the start of program execution.
This ensures that no other memory mappings will block the expansion of the table.
Generic {\em demand paging} is used to do actual memory allocation to support the table as it grows.
This is important as the table {\em cannot be moved} once in use.

For translation, an HTE must contain a pointer to the backing memory of a given handle.
This imparts about eight bytes of overhead per object.
It would be possible to reduce this overhead by compressing the pointer representation as done in prior work \cite{bender2023tiny}, but we have not investigated this potential.
Because all entries of the handle table are the same size, allocation of entries is $O(1)$.
At startup, handle table entries are allocated according to a bump allocator strategy starting from index zero.
When a given HTE is deallocated it is added to a free list for quick reuse by future allocation requests before.
The free list is consulted before bump allocation is used.

While the handle table can always fulfill HTE allocation requests if it has space, it can suffer internal fragmentation from the kernel's perspective.
In the worst case, where the kernel is allocating 2 MiB pages, an adversarial allocation/free pattern could result in a single active HTE per page.
This is unlikely to occur in practice, and active HTE density is quite high in our evaluation.
defragmenting the handle table at the page granularity could be addressed with Mesh.

\subsubsection{The Service Interface}
\label{sec:service-interface}

As described in \secref{sec:design-service-interface}, \alaska{} does not manage the allocation of backing memory itself.
Instead, it defers this task to {\em the service}.
With services, \alaska{} enables a testing ground for future research into the benefits and capabilities of handle-based memory management.
The service interface consists of eight callback functions: two lifetime management functions (\texttt{init}/\texttt{deinit}), two backing memory management functions (\texttt{alloc}/\texttt{free}), and four metadata functions (e.g., query the size of an object).
When the application calls \texttt{halloc}, \alaska{} allocates a handle from the handle table and then requests backing memory from the service via the \texttt{alloc} callback.
The service can later invoke \alaska{} to easily query pin status when moving objects.

\subsection{Battling Fragmentation with \anchorage{}}
\label{sec:anchorage-impl}

The \anchorage{} service uses \alaska{} to implement a defragmenting heap allocator that relies on object movement.

\point{Allocator}
\anchorage{}'s allocator is designed with its freedom to perform defragmentation in mind.
As such, \anchorage{}'s allocator is much simpler than its modern counterparts, which have no choice but to include complex techniques to avoid fragmentation.

\anchorage{} uses a na\"{i}ve bump allocator, and reuses freed memory through the use of a simple power-of-two free-list.
When an allocation is made, this list is searched for an appropriate block in $O(1)$ time (only the front of the list is checked).
If no block is found, allocation is made by bumping a pointer at the top of the heap.
This simple design does not feature the thread caches and other initial-placement optimizations seen in modern allocators.
However, this is merely an engineering limitation.

\anchorage{} subdivides its heap into multiple sub-heaps.
Allocations are made in one sub-heap by first checking the free list before falling back to bump allocation.
When that sub-heap cannot fulfill an allocation request or heap fragmentation is deemed too high, \anchorage{} triggers a {\em barrier} and defragments.

When the runtime barrier fires, all local pin sets are unified.
Unpinned objects are moved from the top of one sub-heap (the source) into another sub-heap (the destination) by simply copying their contents and updating HTEs.
Dictated by a control algorithm (below), \anchorage{} can perform {\em partial defragmentation}---only moving part of the source sub-heap---to amortize the cost of relocating the heap across several pauses.

After each round of defragmentation (both partial and complete) the source sub-heap has as much of its memory returned to the kernel as possible using \texttt{MADV\_DONTNEED}.
This marks the pages of virtual memory as unneeded, and the kernel is free to reclaim them if memory is needed for another process.
With this, resident set size (total physical memory used by a process) increases only momentarily during defragmentation and is quickly reduced.
Unfortunately, invoking the kernel with \texttt{madvise} each round of defragmentation can result in additional TLB shootdowns in multithreaded applications.
A batched technique similar to jemalloc's memory reclamation system could be implemented to reduce this at the cost of additional memory usage \cite{EVANS-JEMALLOC-2015}.

\point{Control system}
\label{sec:defragcontrol}
\anchorage{} can perform a defragmentation pass at any time, but the cost of a pass, and the rate at which they occur constitutes a performance overhead.
In this section, we describe \anchorage{}'s control algorithm to balance the goal of reducing fragmentation with reducing overhead.

This algorithm measures fragmentation using an $O(1)$ metric: the virtual extent of the heap divided by total size of active objects.
The algorithm attempts to keep {\em fragmentation} and the {\em fraction of time spent defragmenting} within bounds set by the operator, $[F_{\mathit{lb}}, F_{\mathit{ub}}]$ and $[O_{\mathit{lb}},O_{\mathit{ub}}]$, respectively.
Both upper and lower bounds are needed to allow for using hysteresis.
The ``aggression parameter'' $\alpha$ controls the fraction of the heap that may be moved during a single pass.

The control algorithm primarily controls $T$, the time to the next defrag event, and operates as a simple state machine.
In the waiting state, the algorithm wakes up every 500ms and checks if the current fragmentation is $>F_{\mathit{ub}}$.
If it is, the algorithm switches to the defragmenting state, where it begins running partial defrag passes, each being limited by $\alpha$.
When a partial defragmentation pass completes, the algorithm measures how long it took, $T_{\mathit{defrag}}$ and controls overhead according to $O_{ub}$ by going to sleep for $T=T_{\mathit{defrag}}/O_{\mathit{ub}}$.
When the algorithm either reaches a fragmentation $<F_{\mathit{lb}}$ or runs out of opportunities, it returns to the waiting state to efficiently observe the system.
A subtle point is that it may not always be possible to achieve the fragmentation bounds, and in this circumstance, we will bounce between waiting and defragmenting quite often, but always stay within $O_{\mathit{ub}}$.

\section{Evaluation} \label{sec:eval}

We now evaluate the efficacy of \alaska{} on a battery of 49 benchmarks from popular suites spanning multiple domains (Embench \cite{bennett2022embench}, GAPBS \cite{beamer2015gap}, NAS 3.0 \cite{bailey1995parallel}, SPEC CPU 2017).
We also test two in-memory databases, Redis and memcached, using the YCSB workload generator \cite{COOPER-YCSB-2010}.
These benchmarks are entirely unmodified, together constituting nearly 3 million lines of code.
With this evaluation, we seek to answer the following questions:

\begin{enumerate}[label={\bf Q\arabic*}, leftmargin=*]
	\item Does the system truly deliver automatic transparent handles? How much programmer effort is involved?
  \item What effects do \alaska{}'s translations and tracking have on code size?
	\item What is the software engineering effort required to build the \alaska{} system?
	\item What is the performance overhead of software handles?
	\item How effective is \alaska{} at enabling \anchorage{} to reduce fragmentation in a real-world application?
  \item What is the effect of \anchorage{}'s stop-the-world pause times in a multithreaded application?
\end{enumerate}

\subsection{Experimental Setup}
\label{sec:eval-setup}

Our evaluation was performed on a Dell R6515 with a single AMD EPYC 7443P running Ubuntu 22.04.1 LTS.
Each processor has 24 cores with 2-way SMT, 64 entry store buffer, 768 KiB L1D\$, 12MiB L2\$, and 128MiB L3\$ all with 64B line size backed by 512 GiB of DDR4 RAM at 3200 MT/s.
Our implementation of \alaska{} is built atop LLVM 15.0.1.
All performance results are gathered from 10 executions per configuration, and speedup/overhead metrics are relative to the median of baseline execution.
Each compilation is performed by first applying the \texttt{-O3} optimization level with OpenMP disabled.
\alaska{} transformations are then applied to the LLVM bitcode, followed by inlining and several optimization passes.
Those passes are similarly applied to the baseline bitcode to ensure a fair playing field.

\subsection{\alaska{} Achieves Programmer Transparency}
\label{sec:eval-transparency}

To evaluate \alaska{}'s ability to provide fully automatic, transparent handle-based memory management, we recorded how many edge cases we had to handle in developing the system.
Throughout development, we constantly tested \alaska{} against both Redis and all of the aforementioned benchmarks.
Outside of the issues solved by escape handling and \texttt{musl} substitutions (\secref{sec:compiler-escape-impl}), {\bf \em no code in any benchmark or application was modified to support handles.}
Making these benchmarks handle-aware was entirely achieved outside of the application code, by the \alaska{} compiler and runtime.

Unfortunately, some benchmarks---namely perlbench and gcc from SPEC ---violate the strict aliasing assumptions (\secref{sec:assumptions}) of \alaska{} .
This is not a fundamental limitation of the handle approach, but rather a limitation of our implementation's hoisting technique in the compiler, which relies on the LLVM memory model.
If the \texttt{-fno-strict-aliasing} flag is passed to the compiler, both of these benchmarks {\em function correctly}.
This flag instructs \alaska{} to disable the hoisting optimization, instead translating handles before each load and store.

Given that no benchmark or application was modified to support \alaska, we argue that the answer to {\bf Q1} is yes.
Most applications, {\em including Redis}, can be transformed to use automatic transparent handles with a single command: \\
\centerline{
\texttt{make CC=alaska CXX=alaska++}
}

\textbf{Q2} concerns itself with compilation overhead.
\alaska{}'s compiler does not significantly increase compilation time.
Redis takes an additional 12 seconds to compile with \alaska{} (up 21\% from 57 seconds, single-threaded), however, this is due to a relatively unoptimized compiler implementation.
Most of this extra time comes from the requirement to compile the whole program to correctly handle calling library code.
Optimizing \alaska{}'s link-time transformations \`a la ThinLTO \cite{THINLTO-JOHNSON-2017} or Propeller \cite{SHEN:PROPELLER:2023} could reduce this overhead.

Executable files grew only about 48\% (geomean) via the \alaska{} transformations and runtime system.
The worst case is a doubling of executable file size which occurs when the hoisting optimization is not applicable.
For example, \verb|xalancbmk| doubles in size as it makes use of linked lists, which have short handle translation lifetimes.
Applications from NAS, which benefit heavily from hoisting, see negligible increases in code size.

\subsection{Building and Extending \alaska{} is Practical} \label{sec:eval-se}

We inspect the code size and development time to answer \textbf{Q3}.
\alaska{} is not a particularly large codebase, and was built over a period of eight months.
Its core runtime is only 1316 lines of C++, which manages structures like the handle table, tracking, and signaling for barriers.
The largest effort was the compiler, which constitutes 2393 lines of C++.
The \anchorage{} service is a mere 567 lines of C++, {\em half the size of the bespoke defragmentation system in Redis} \cite{REDIS-ACTIVEDEFRAG}.
\anchorage{} is general purpose and does not leak into application code, unlike activedefrag.
We take away that, while \alaska{}'s core may have been complex to design and implement, extending it with additional services is trivial, and we have already begun experimenting with future research directions.

\subsection{Overhead Varies, but is Overall Low} \label{sec:eval-overhead}

\begin{figure*}[th]
	\includegraphics[width=1\linewidth]{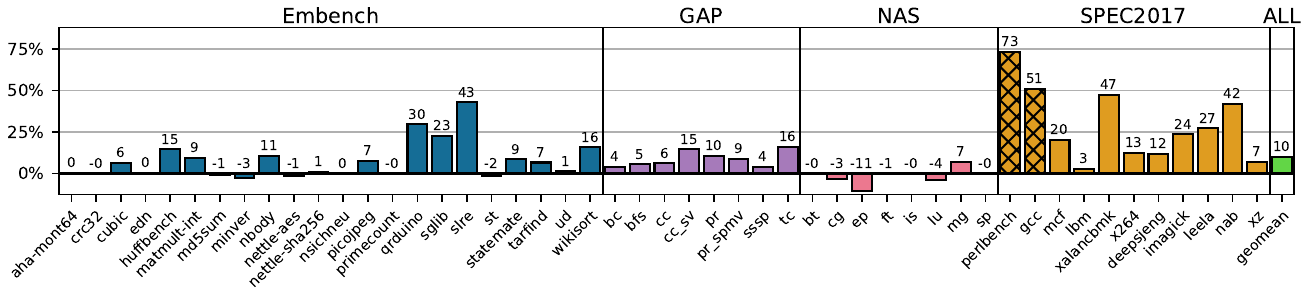}
	\caption{
    The \alaska{} prototypes' overhead (\% increase) from translation and pin tracking is about 10\% with several outliers
	}
	\label{fig:alaska-overhead}
\end{figure*}

To answer \textbf{Q4}, we evaluated \alaska{}'s runtime overhead on benchmarks from several domains.
\figref{fig:alaska-overhead} compares the measured performance overhead (percent increase of wall-clock time) of the benchmarks running \alaska{} without a service (using malloc to allocate backing memory) against a baseline described in \secref{sec:eval-setup}.
The Perlbench and GCC benchmarks from SPEC CPU 2017 both violate the assumptions listed in \secref{sec:assumptions}, and have been compiled with hoisting disabled.
The main takeaway is that \alaska{}'s performance impact varies depending on the benchmark, but is overall relatively low, imparting a geomean of 10\% overhead if perlbench and GCC are included, and an 8\% geomean if not.

Despite the outliers, many benchmarks have {\em near zero overhead}.
These benchmarks benefit substantially from the hoisting capability of \alaska{}'s compiler, with many benchmarks having translations hoisted to their outermost loops.

For example, \texttt{619.lbm\_s} from SPEC CPU features a large grid allocation which is accessed inside a series of inner loops.
The translations in this program are all successfully hoisted to their outermost possible loop level, and as a result, the cost of \alaska{} is amortized to a high degree.
Similar amortization can be seen in the NAS benchmarks and \texttt{xz} from SPEC CPU 2017, which feature similar structures in performance-sensitive parts of the application.

Unfortunately, not all programs feature this friendly structure, and some programs do not benefit from any of \alaska{}'s optimizations.
These benchmarks stress \alaska{}'s ability to hoist and efficiently translate pointer-chasing data structures such as linked lists and trees for which the compiler is unable to amortize the cost of translation.
Additionally, several benchmarks in the Embench suite suffer from poor software design patterns, which block any hoisting from occurring.
One such is in \texttt{sglib}, which passes all parameters to the core kernel through {\em global variables} instead of as arguments.
In LLVM, this results in an additional load from the global variable every time it is used, and hoisting the translation across these may break correctness in concurrent applications.

Handle translation overhead is also seen in benchmarks that perform very little work per translation, such as \texttt{sglib}, \texttt{mcf} and \texttt{xalancbmk}.
For example, the \texttt{mcf} benchmark from SPEC CPU spends roughly 40\% of runtime sorting an array of pointers, which results in 4 translations per comparison.
Similarly, \texttt{xalancbmk} is written using C++ virtual methods, which block almost all optimizations that would optimize across call sites, and the \texttt{this} pointer is translated in almost every function---even when it is not required.

Interestingly, we see a {\em speedup} of 11\% in NAS ep.
In this benchmark, we see an increase in L1 instruction cache accesses and misses.
The most probable cause of this is the differences in code layout by the LLVM backend that we observed.
The effect of this is exacerbated by how small the application is, with the hot loop being only 256 bytes.

\begin{figure}
	\includegraphics[width=1\linewidth]{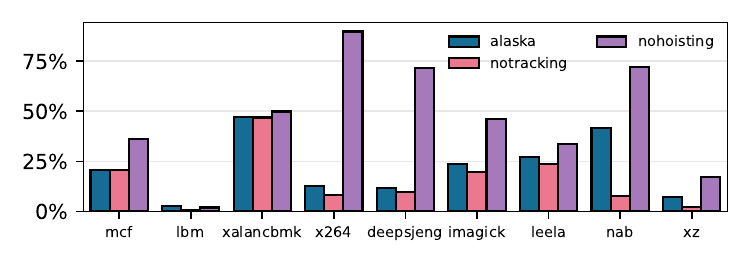}
  \caption{The hoisting optimization drastically reduces the overhead of \alaska{} handles, and careful design and engineering effort of the tracking system imposes negligible overhead over the cost of using handles with the exception of a few applications (nab, xz)}
  \label{fig:ablation}
\end{figure} 

To investigate the source of the overheads in these benchmarks, we ran an ablation study on the SPEC CPU 2017 benchmarks where we removed features of \alaska{} and evaluated the resulting overhead.
The performance overhead results of this study are shown in \figref{fig:ablation}.
The metric marked ``alaska'' is the overhead seen in \figref{fig:alaska-overhead} with both hoisting and tracking enabled.
When the compiler's hoisting optimization is disabled, as seen in the ``nohoisting'' metric, most benchmarks see their overhead nearly double.
Those benchmarks that do not see a large increase---such as xalancbmk---do not have many opportunities to hoist.

When the tracking system is removed, marked ``notracking'', several benchmarks see a large reduction in overhead.
This is most apparent in \texttt{nab} and \texttt{xz}, and we attribute this to the experimental nature of the LLVM StackMap system, which has seen very little use outside of JIT compilers. This overhead mostly comes from the insertion of poll points (\secref{sec:tracking-impl}), which in the common case should incur no additional overhead---the polls are simple \texttt{NOP} instructions.
However, it is possible that either the addition of these instructions causes undue stress to the instruction cache or, more likely, this experimental system leads to unexpected {\em performance} bugs in LLVM's backend for some application structures.

\alaska{}'s overheads could be further improved with memory analysis \cite{steensgaard1996points,sui2016svf,GUO:VLLPA:2005,MCMICHEN:2024:MEMOIR}, which can say when a value is {\em definitely} a handle or not.
This would allow the compiler to completely eliminate the conditional check and branch before translation.
Historically, the removal of these checks from read barriers have reduced overheads significantly \cite{BLACKBURN:BARRIERS:2004}.

\subsection{\anchorage{} Defragments without Black Magic}
\label{sec:redis-defrag}
To evaluate the capabilities of the \anchorage{} service built on \alaska{}, we tested it using a large, unmodified application: Redis \cite{REDIS-WEBSITE}, an extremely popular key-value in-memory database.
As we described in the introduction, Redis has long suffered from fragmentation, particularly due to its tendency to be used as an LRU cache atop other databases in large deployments, resulting in a heap that has allocations scattered everywhere as old objects are freed to make space for new ones.
As a result, Redis includes activedefrag, a bespoke defragmentation system requiring {\em manual modifications throughout the application's code}, which is also considered ``black magic'' \footnote{A term used by the original developer:\\ \url{https://twitter.com/antirez/status/1052590584102305792}} and cannot be reused in other applications.

\point{Response latency} We begin by considering the impact on response latency.
We drive Redis with the YCSB workload generator using two workloads \cite{COOPER-YCSB-2010}.
We found that \anchorage{} on top of \alaska{} imparts an average of 13\% overhead on read latencies (Workload A), and an average 17\% overhead on update/write latencies (Workload F).
Aside from the overhead of \alaska{} discussed in \secref{sec:eval-overhead}, the lower throughput of the \anchorage{} allocator imparts an additional overhead relative to glibc malloc used by the baseline.
Of note, the allocator overhead is not intrinsic and could be ameliorated through improvements and optimizations in its design.

\point{Defragmentation} To answer \textbf{Q5}, we used the same workload from Mesh \cite{POWERS-MESH-2019} which configures Redis to limit memory usage to 100 MiB and then inserts more than that into the database using a workload generator.
Redis then evicts keys from its dataset using an LRU policy until memory usage falls below the 100 MiB threshold.
This creates significant fragmentation, as there are holes throughout the heap that are not filled---and the application's RSS does not decrease in the baseline allocator.
Configuring Redis in this way is very common when it is used as a cache for other datasets.
This benchmark runs for a total of 10 seconds and includes results for the baseline allocator, \anchorage{}, \texttt{activedefrag}, and Mesh.

\begin{figure}[t]
	\includegraphics[width=1\linewidth]{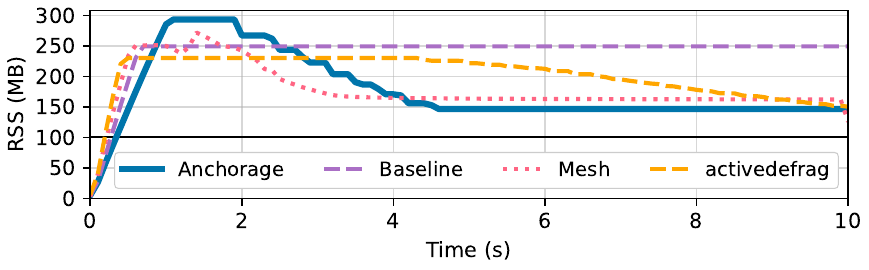}
  \caption{\anchorage{} can defragment memory under Redis at least as well as the bespoke Redis defrag implementation.}
  \label{fig:defrag-redis}
\end{figure} 

\figref{fig:defrag-redis} shows the results.
\anchorage{} effectively reduces memory usage from almost 300MiB to 150MiB (40\% less compared to baseline Redis).
The main takeaway is that \anchorage{}, which is readily applicable to any codebase without any code modifications, is able to do as well as \texttt{activedefrag}, which requires extensive, manual, hand-rolled changes to the codebase.
We include the data verbatim from Mesh.

\begin{figure}[t]
	\includegraphics[width=1\linewidth]{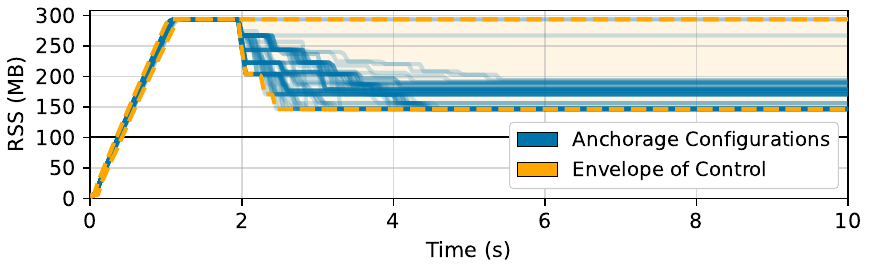}
  \caption{\anchorage{} gives users control over the tradeoff between overhead and fragmentation.}
  \label{fig:defrag-redis-control}
\end{figure} 

\point{Control} To evaluate the effect of our control system, we sweep the parameters $[F_{\mathit{lb}}, F_{\mathit{ub}}]$, $[O_{\mathit{lb}},O_{\mathit{ub}}]$, and $\alpha$ (\secref{sec:defragcontrol}).
In \figref{fig:defrag-redis-control}, each solid curve shows \anchorage{}'s behavior with a different parameter set, of which there are hundreds.
Each parameter set stays within the overhead bounds ($[O_{\mathit{lb}},O_{\mathit{ub}}]$).
As shown in the figure, the envelope of control (dashed curves) is quite large, meaning that the parameters have a significant effect.
In a deployment, the user can tune these parameters (dynamically, even) for their desired tradeoff between overhead and fragmentation.

\begin{figure}[t]
	\includegraphics[width=1\linewidth]{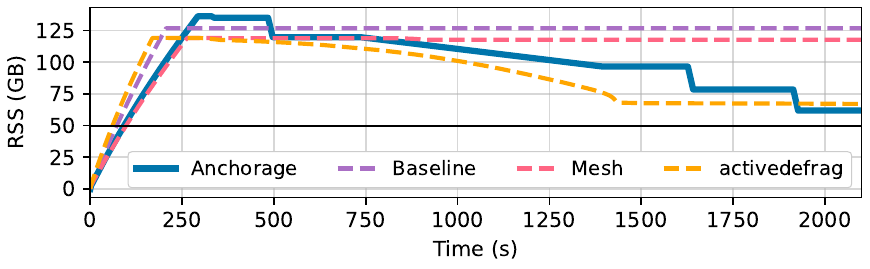}
  \caption{\anchorage{} successfully defragments workloads with $>$100GiB RSS.}
  \label{fig:defrag-redis-big}
\end{figure}

\point{Stressing \anchorage{}}
To evaluate \anchorage{} in the face of workloads with a large amount of memory we adapted the test from \figref{fig:defrag-redis} to have a maximum memory policy of 50 GiB, instead of the original 100 MiB.
This workload inserts 100 GiB of data, 500 bytes at a time, causing over 2.5x fragmentation--Redis' internal datastructures provide some overhead--when eviction begins around 250 seconds.
\figref{fig:defrag-redis-big} shows the results of this experiment with \anchorage{} handily defragmenting the heap of this large workload, achieving similar steady-state RSS as \texttt{activedefrag}, albeit over a longer time frame.

The longer time taken by \anchorage{} to defragment the heap is caused by its control algorithm.
Around 500 seconds into the test, the control algorithm begins defragmentation.
Because the control system operates in units of \textit{percentage of the heap}, the system immediately enters a \textit{7 second pause} as it drastically mispredicts how long the defragmentation will take.
The system then backs off for over 250 seconds to maintain the 5\% overhead maximum---per its configuration---and begins slowly defragmenting the heap for the rest of the application runtime.
It is important to note that this is not a fundamental issue of \anchorage{}, as the control system can be tuned---as illustrated in \figref{fig:defrag-redis-control}.
As such, careful parameterization of the control algorithm could aid in preventing this behavior.
Alternatively, a common approach to hide GC pause times is via concurrent algorithms, the potential of which is briefly discussed in \secref{sec:discussion}.

We also evaluated Mesh in this environment and, in its default configuration, it was either not aggressive enough---defragmenting only a few MiB at a time as can be seen around 750 seconds---or does not scale to workloads this large.
Note, we had to modify Mesh's source code to allow heap sizes larger than 64GiB.

\subsection{\anchorage{}'s Stop-the-World Pauses}
\begin{figure}[t]
	\includegraphics[width=1\linewidth]{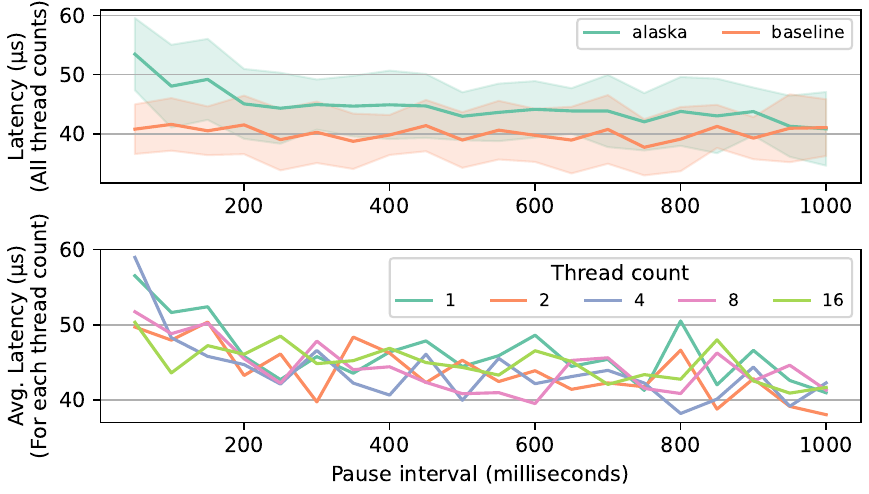}
  \caption{Latencies in \texttt{memcached} vary based on the pause time, and there is no trend between latency and number of threads.}
	\label{fig:memcached-tput}
\end{figure}

To answer \textbf{Q6}, we evaluated \anchorage{}'s effect on request latencies on an alternative in-memory key-value database, \texttt{memcached} \cite{MEMCACHED-WEBSITE}, which can be configured to run in parallel with multiple threads.
We devised a synthetic test where $\sim$1 MiB of memory is relocated at each pause, regardless of the fragmentation ratio, resulting in average pause times less than 2ms.
\texttt{memcached} is driven by the YCSB test suite---specifically workload A, which has been scaled up to run for longer---which provides the latencies shown.
Because this workload is driven through the loopback network, it does see considerable noise.
The results of this test are shown in \figref{fig:memcached-tput}, gathered by varying the number of threads and the interval at which \anchorage{} performs a pause.

The upper plot in \figref{fig:memcached-tput} shows the overall effect on latency for all thread configurations combined, as well as the standard deviation, indicating that \alaska{} incurs an average of 10\% overhead in this application across all configurations (including impractical 50ms pause intervals).
This manifests as an average latency increase of $4\mu s$.
With more practical intervals (above 500ms), the latency increase drops to less than 7\%.
The lower plot breaks the latencies down by thread count.
The main takeaway is that, at low intervals, there is an expected effect on average latency.
This overhead is driven primarily by outliers when requests are blocked by \alaska{} runtime pauses.
Importantly, \anchorage{} never naturally pauses at these low intervals in other applications.
Additionally, we measured no correlation between number of threads and pause time.

\section{Related Work} \label{sec:related}

\point{Defragmenting Unmanaged Languages} has seen a resurgence in recent years.
 Mesh \cite{POWERS-MESH-2019}, the most pertinent example,
works on existing binaries without recompilation.  Mesh changes
malloc so that heap objects are carefully positioned on virtual pages.
A cooperating kernel takes advantage of this by mapping several virtual pages to the same physical one without
overlapping the heap objects, reducing fragmentation and memory usage at the
physical level.  In contrast, \alaska{} enables heap object mobility in the virtual address space,
a more general capability that is necessary to achieve
fragmentation limits \cite{ROBSON-1977}, and can enable other
mobility-enabled services beyond defragmentation.  

\point{Compiler-managed Address Translation}
CARAT \cite{CARAT-SUCHY-2020,CARATCAKE-SUCHY-2022} aims to replace
paging via per-object RWX protections and object-level movement in the
physical address space.  Both functionalities are enabled by
compiler/kernel cooperation and are automatic and transparent to the
programmer.  Object mobility is based on allocation and pointer escape
tracking, combined with updating of all pointers.  In other words, there
is no indirection like with handles in \alaska{}.  This avoids the
handle pin and handle translation costs, but makes object movement
more expensive because the update is not $O(1)$ as in handles, but
at least $O(m)$, where $m$ is the number of pointers to the object.  While
\alaska{} does not currently enable protections, it could do so in the
future.

\point{Programming Models}
Interestingly, the historic manual indirection
strategies described in \secref{sec:handles} have returned.
AIFM \cite{RUAN:2020:AIFM} features a programming model on
top of C++ to allow the developer to take advantage of far-memory
data structures.  While the system works well in enabling far memory,
it requires the programmer to do error-prone heavy lifting, insert
scopes, and pinning logic similar to classic handle-based systems.
Recent work to automatically transition an application to a far-memory application through compiler techniques such as TrackFM \cite{TAURO:TRACKFM:2024} or Mira \cite{GUO:MIRA:2023} are promising.
Internally, these automatic systems utilize many of the techniques used by \alaska{}, such as the level of indirection and compiler hoisting.
We suspect that \alaska{}'s service abstractions may be sufficient to implement these systems.

Oilpan \cite{OILPAN}, a library that enables precise garbage collection in C++ programs boasts limited compaction capabilities, but requires the application be rewritten to use Oilpan abstractions.
Similarly, ActivePointers and others \cite{SHADAR:2016:ACTIVEPOINTERS,Detlefs1992GarbageCA,EDELSON:1992:CPPGC}
allow the runtime to relocate objects and invalidate references (a la
paging) via a modification of C++'s ``smart pointers''. These
approaches demand rewriting parts of the application or
using libraries that cannot benefit from translation-aware
optimizations in the compiler.  Such systems could perhaps be
easily built atop \alaska{}'s service interface.

\section{Discussion} \label{sec:discussion}

Though we have only discussed handles in the context of fragmentation, we note that handles provide the more fundamental capability of managing object motion at the object granularity.
Handles in the past have also given the ability to move or swap memory at the object granularity, {\em rather than pages}.
Managing memory at the object granularity has gained interest in several contexts, such as application-level remote memory systems \cite{RUAN:2020:AIFM}, replacements for paging \cite{CARAT-SUCHY-2020,CARATCAKE-SUCHY-2022}, and security via capability-based addressing \cite{CHERI-WOODRUFF-2014,CHERI-WATSON-2015}.
It has also been shown that object mobility can be used to dynamically enhance cache locality \cite{CACHE-CONSCIOUS-CHILIMBI-1999,COURTS-1988,DATA-REORG-DING-1999,AUTO-RESTRUCTURE-LINKED-SPEK-2010,ON-THE-FLY-SPLITTING-WANG-2012}.
Similarly, work has been done to place rarely used objects in cheaper nonvolatile memory to optimize memory usage \cite{MERCHANDISER}.

We posit that handles provide a powerful vehicle to implement such ideas with simple modifications to the compiler and runtime.
The \alaska{} system has been designed to be extensible beyond what we describe in this paper with these capabilities in mind.
\alaska{} can be configured with ``handle faults'', which very closely approximates the capabilities of a system that uses page faults.
While this paper does not evaluate this feature in detail, initial investigation indicates that this check has minimal additional overhead (\textasciitilde 1-2\%), but enables advanced techniques in the runtime system.
With this check enabled, objects can be swapped out of memory in the same way a kernel might do so for pages.

Further, this ``swapping'' mechanism could be utilized to {\em speculatively} move memory without stopping the world for the duration of movement.
The runtime would occasionally mark entries in the handle table as ``invalid'' and speculatively copy their data to an alternative location.
If another thread accesses that handle, it would trap to the runtime and atomically mark the object as ``valid''.
The runtime would then attempt to CAS (compare-and-swap) the entry in the handle table, marking it as ``valid'' with a new address.
If the CAS succeeds, the old memory can be freed, and the object has been relocated.
If it fails, the relocation is aborted, and the speculative copy is freed, as some other thread has pinned that handle while the copy was being made.
We see it as an interesting path forward to implement concurrent memory movement, as this closely resembles the concurrent compaction system seen in the Shenandoah GC \cite{SHENANDOAH}.
This simple mechanism could be utilized to implement swapping objects to disk, compression, or even far memory.

\section{Conclusion} \label{sec:conclusion}

We have described the design and implementation of \alaska{}, a
prototype compiler and runtime system that automates the use of
handle-based memory management while being entirely transparent to the
programmer. \alaska{} is a platform that enables runtime features and
services that rely on object mobility.  Using this platform, we
designed and implemented \anchorage{}, a defragmenting memory allocator
that reduces memory usage in highly fragmented applications.
In the future, we plan on using \alaska{}'s extensibility as a vessel
to enable additional transparent runtime services such as memory
disaggregation, locality enhancement, and capability-based security,
as well as a lightweight alternative to paging.

\subsection*{Acknowledgments}
We thank the anonymous reviewers as well as our shepherd Steve Blackburn for their time and feedback.
We also thank members of the Prescience and ARCANA Labs for their support and feedback on this work.
This effort is based upon work supported by the U.S. National Science Foundation (NSF) under awards CCF-2119068, CNS-2211315, CNS-1763743, CCF-2028851, CCF-2107042, and CCF-1908488.
This project was also supported by the United States Department of Energy via the grant DESC0022268.

\newpage

\appendix
\section{Artifact Appendix}

\subsection{Abstract}

Our artifact includes the source code for the \alaska{} prototype, as well as tooling to automatically test it against a bevy of benchmarks and applications.
The output of this artifact is the data required to generate the paper's figures, as well as the figures themselves.
This artifact can be optionally used to evaluate \alaska{}'s overheads in the SPEC CPU 2017 benchmark suite, as outlined in \secref{sec:softdep}.
The artifact also features an in-depth \verb.README. file, which includes example results from the paper, as well as directions on how to use \alaska{} in other C/C++ aplications.

\subsection{Artifact check-list (meta-information)}

{\small
\begin{itemize}
  \item {\bf Program:  The \alaska{} compiler and runtime. }
  \item {\bf Compilation: Clang+LLVM is leveraged as the basis for the \alaska{} compiler transformations. These are downloaded automatically. }
  \item {\bf Transformations: The \alaska{} compiler, including hoisting and tracking optimizations. }
  \item {\bf Run-time environment: The \anchorage{} defragmenting allocator. }
  \item {\bf Output: Figures 7, 8, 9, 10, 11, and 12.}
  \item {\bf Experiments: Overhead on Embench, GAPBS, NAS and SPEC CPU. Memory defragmentation for Redis, and throughput measurements of Memcached. }
  \item {\bf How much disk space required: 32GB}
  \item {\bf How much time is needed to complete experiments (approximately)?  On our hardware, 24-48 hours are needed if SPEC CPU is evaluated, \textasciitilde 7 hours if not.}
  \item {\bf Publicly available? Yes. }
  \item {\bf Code licenses (if publicly available)? MIT License. }
  \item {\bf Workflow framework used? Docker. }
  \item {\bf Archived (provide DOI)?  Will be created for the final appendix}
\end{itemize}
}

\subsection{Description}

\subsubsection{How to access}

The artifact can be accessed from Github at \url{https://github.com/PrescienceLab/alaska-asplos24-artifact}.
The DOI for the artifact is 10.5281/zenodo.6350453 on zenodo.
The repo also features an informative \verb.README. for using the artifact.

\subsubsection{Hardware dependencies}

An x86\_64 system with eight or more cores, 32GB+ of memory, and more than 32GB of free disk space is required.
If you wish to generate Figure 11, a machine with at least 200GB of memory is required.  
Our testbed features an AMD EPYC 7443P with 512GB of memory.

\subsubsection{Software dependencies}
\label{sec:softdep}
The artifact has been extensively tested on, and is designed for, an Ubuntu 22.04 system.
The \verb.README. lists dependencies as they can be installed from \verb.apt., and all other software such as LLVM are downloaded automatically by the artifact's test harness.
We recommend running in a Docker container, which is provided.

The artifact evaluates four standard benchmark suites.
The three open source suites (NAS, Embench, and GAPBS) are downloaded automatically.
SPEC2017 can be optionally evaluated if it is available to the user of the artifact (i.e., if the user has a license).    
The artifact functions correctly without SPEC CPU.  If SPEC CPU is unavailable it will simply not include SPEC CPU benchmarks in Figure 7, and will skip producing Figure 8 alltogether.
Directions regarding SPEC CPU can be found in either the \verb.README. or below.

\subsection{Installation}
Once the artifact repo is downloaded the first (optional) step is to place the SPEC2017 source tarball in the root of the repository.  It must be named \verb|SPEC2017.tar.gz|.
If found, the test harness will extract and compile it.
If not, the test harness will simply not evaluate SPEC CPU.

Because runs can take a long time, we recommend starting a \verb|tmux| session to avoid having disconnections disrupt runs.

The docker container can be started with \verb|make in-docker| (or \verb|make in-podman|, if that is preferred).
This will start a bash shell in an ephemeral docker container, in which the directory \verb|/artifact| is bind\-mounted to the host filesystem.
Any changes in this container will be reflected in the host filesystem automatically.

\subsection{Experiment workflow}

This artifact features a fully automatic workflow to generate all the experimental results included in the paper.
It compiles \alaska{}, compiles and runs benchmarks, produces data, and plots the results automatically.

The results of compilation can be found at the top level of the repo in \verb|./opt/|, which contains several enable scripts which can be utilized to use \alaska{} (more on this in the \verb.README.).   

\subsection{Evaluation and expected results}

After downloading the repo and following the installation instructions, simply run \verb|./run_all.sh| in the top level of the repo.  
You will be prompted with several yes/no questions.
An important one is whether you wish to generate Figure 11, which requires $>250$GB of memory.
If your test machine does not feature that much memory, answer no, and that test will be skipped.

Additionally, \verb!run_all.sh! will search for the SPEC 2017 tarball that was optionally included in the installation phase.
If it was not found, the script will notify you of this, and you can chose to continue without SPEC CPU, or exit.

Once the run is finished, which can take many hours, the \verb.results/. directory will be populated with data in the form of csv files, as well as a series of figure PDFs, which correspond to the identically numbered figures in the paper:

\textbf{results/figure7.pdf}:
Evaluates the overhead of \alaska{}'s handles on a series of benchmarks.
If SPEC CPU was not included, it will not be included.

\textbf{results/figure8.pdf}:
Evaluates the effectiveness of \alaska{}'s compiler optimizations.
If SPEC CPU was not included, this figure will not be generated.

\textbf{results/figure9.pdf}:
Evaluates the effectiveness of \anchorage{}'s defragmentation capabilities.
This test includes results of \anchorage{}, activedefrag, and Mesh.

\textbf{results/figure10.pdf}:
Evaluates that \anchorage{} can be configured in many ways, resulting in many different rates of defragmentation.
The this test uses randomized configurations, so the figure may not match exactly.

\textbf{results/figure11.pdf}:
This figure tests how alaska manages to defragment large\-memory workloads. 
This test is very similar to Figure 9, except it allocates significantly more memory.

\textbf{results/figure12.pdf}:
This figure shows an evaluation of \alaska{}'s effect on multithreaded applications, using memcached with varying thread counts.

More details on these figures can be found in the arifact's \verb|README.md| file.

If there are any problems with the artifact, first try \verb|make| \verb|distclean| to reset the repo.

\subsection{Notes}

We assume an internet connection through the duration of the benchmarking phase, and that the machine running the artifact is not running an existing copy of either Redis or Memcached, as we configure these to run on their default ports.


%
%


\bibliographystyle{plainurl}
\balance
\bibliography{paper}

\end{document}